\documentstyle[aps,prb,preprint,epsfig]{revtex}
\draft
\begin{document}

\title{
 Effects of nonmagnetic impurities on optical conductivity
 in strongly correlated systems
}
\author{Seung-Pyo Hong, Kwangyl Park, and Sung-Ho Suck Salk}
\address{
 Department of Physics, 
 Pohang University of Science and Technology \\
 Pohang 790-784, Republic of Korea
}
\maketitle

\begin{abstract}
Effects of nonmagnetic impurities
on optical conductivity in the systems of 
antiferromagnetically correlated electrons
are examined based on the Lanczos exact diagonalization scheme.
As a result of resonant scattering
a low-frequency peak in the optical conductivity
is predicted to occur in the presence of the nonmagnetic impurities,
which is consistent with the observed normal-state optical conductivity
of ${\rm YBa_2}({\rm Cu}_{1-x}{\rm Zn}_x)_3{\rm O}_{7-\delta}$.
In addition, a relatively high and broad peak is found to occur
at a high frequency region
as a consequence of the Heisenberg interaction between electrons,
in agreement with observation in the peak position.
\end{abstract}

\vspace*{1cm}


\newpage

\section{Introduction}
Optical spectroscopy is an important tool
in probing electronic states of strongly correlated systems
including high-$T_c$ cuprates.
Infrared properties of optical conductivity
for ${\rm YBa_2Cu_3O}_{7-\delta}$ crystals
have been intensively investigated 
to explore the low-energy dynamics of charge carriers.
\cite{88Timusk,89Cooper,90Orenstein}
Optical conductivity
of ${\rm La}_{2-x}{\rm Sr}_x{\rm CuO_4}$ 
for several doping rates between $x=0$ and $x=0.34$ at room temperature
exhibited the appearance of mid-infrared band near 0.5 eV
with the increase of hole doping concentration.\cite{91Uchida}
The mid-infrared band was also observed 
for nonstoichiometric cuprates
of ${\rm Nd_2CuO}_{4-y}$ and ${\rm La_2CuO}_{4+y}$
with some vacancies on oxygen sites.\cite{92Thomas}
Exact diagonalization calculations on small clusters
verified the existence of the mid-infrared band
upon doping holes away from half filling.
\cite{90Moreo,92Dagotto,94Dagotto}

Nonmagnetic impurities embedded in the high-$T_c$ cuprates
have been used to investigate transport and magnetic
properties of the cuprates.
A small amount of Zn substituted for Cu is known to appreciably reduce
the superconducting transition temperature.\cite{90Xiao}
Magnetic susceptibility data\cite{90Xiao} for 
${\rm La_{1.85}Sr_{0.15}Cu}_{1-x} {\rm Zn}_x {\rm O_4}$
and NMR data\cite{91Alloul} for
${\rm YBa_2} ({\rm Cu}_{1-x} {\rm Zn}_x)_3 {\rm O_{7-\delta}}$
provided evidences that Zn induces
magnetic moments in the ${\rm CuO_2}$ plane.
However, not much attention has been paid to 
the effect of nonmagnetic impurities on optical conductivity
of the cuprates.\cite{98Wang}
In the present paper we report
an exact diagonalization study of optical conductivity
when a nonmagnetic impurity is introduced into 
the systems of antiferromagnetically correlated electrons.

\section{Optical Conductivity}
We consider the following model Hamiltonian
for the study of optical conductivity
in strongly correlated electron systems,
\begin{equation}
H = -t \sum'_{\langle ij \rangle \sigma}
    (\tilde{c}_{i\sigma}^\dagger \tilde{c}_{j\sigma} + \mbox{H.c.})
  + J \sum'_{\langle ij \rangle} \left( {\bf S}_i \cdot {\bf S}_j  
    - \frac{1}{4} n_i n_j \right)
  + V_{\rm imp} \sum_{\ell} (1-n_\ell) ~.
\label{eq:ham}
\end{equation}
Here $\tilde{c}_{i\sigma}^{}$
is the electron annihilation operator
at site $i$ with no double occupancy.
${\bf S}_i = \frac{1}{2} c_{i\alpha}^\dag \bbox{\sigma}_{\alpha\beta}^{}
c_{i\beta}^{}$ is the electron spin operator
and $n_{i}^{}=\sum_\sigma c_{i\sigma}^\dag c_{i\sigma}^{}$
is the number operator.
$t$ is the hopping energy
and $J$, the Heisenberg exchange energy.
The prime in $\sum'_{\langle ij \rangle}$ denotes 
the sum over the nearest neighbor links $\langle ij \rangle$
only between copper sites, thus excluding
the impurity site.\cite{94Poilblanc}
Nonmagnetic impurities Zn substituted for Cu atoms have
closed-shell configuration of ${\rm Zn^{2+}}$ $(3d^{10})$
and are inert to electron hopping.
The one-body Coulomb potential of the impurity,
$V_{\rm imp}$ represents 
a repulsive interaction as a result of
a positive charge (${\rm Zn^{2+}}$ ion)
of the nonmagnetic impurity
interacting with a doped charge carrier (hole).
$\sum_\ell$ is the sum over the nearest neighbor links
with the impurity site.

The optical conductivity is obtained from\cite{90Moreo}
\begin{equation}
\sigma(\omega) = - \frac{1}{\omega\pi} {\rm Im}
 \left\langle\psi_0 \left| j_x \frac{1}{\omega-H+E_0+i\epsilon}
 j_x \right|\psi_0 \right\rangle ~,
\end{equation}
where $j_x$ is the current operator in the $x$-direction,
\begin{equation}
j_x = it \sum_{{\bf i},\sigma}
 (c_{{\bf i+x}\sigma}^\dag c_{{\bf i}\sigma}^{}
 - c_{{\bf i}\sigma}^\dag c_{{\bf i+x}\sigma}^{})
\end{equation}
and $|\psi_0\rangle$ is the ground state of energy $E_0$.
For the study of optical conductivity
in the presence of impurity for a hole-doped system  
we introduce a nonmagnetic impurity atom and one mobile hole
into the systems of
$4\times4$ and $\sqrt{20}\times\sqrt{20}$ square lattices
with periodic boundary conditions,
in order to allow Lanczos exact diagonalization calculations.

\begin{figure}[h]
\vspace*{20mm}
\centering
\epsfig{file=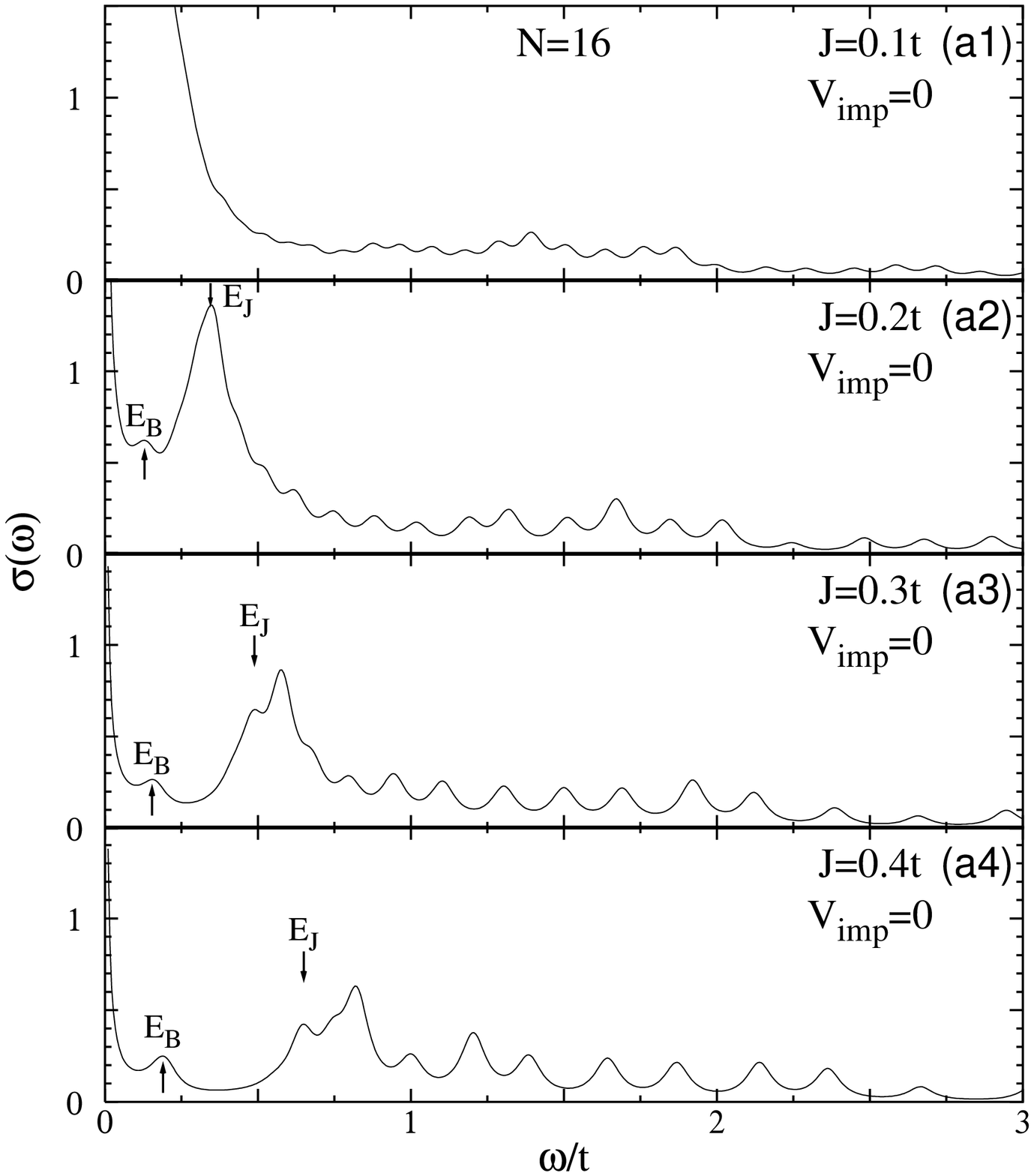, width=12cm} \\
(a) \\
FIG.~1. \\
\newpage
\vspace*{24mm}
\epsfig{file=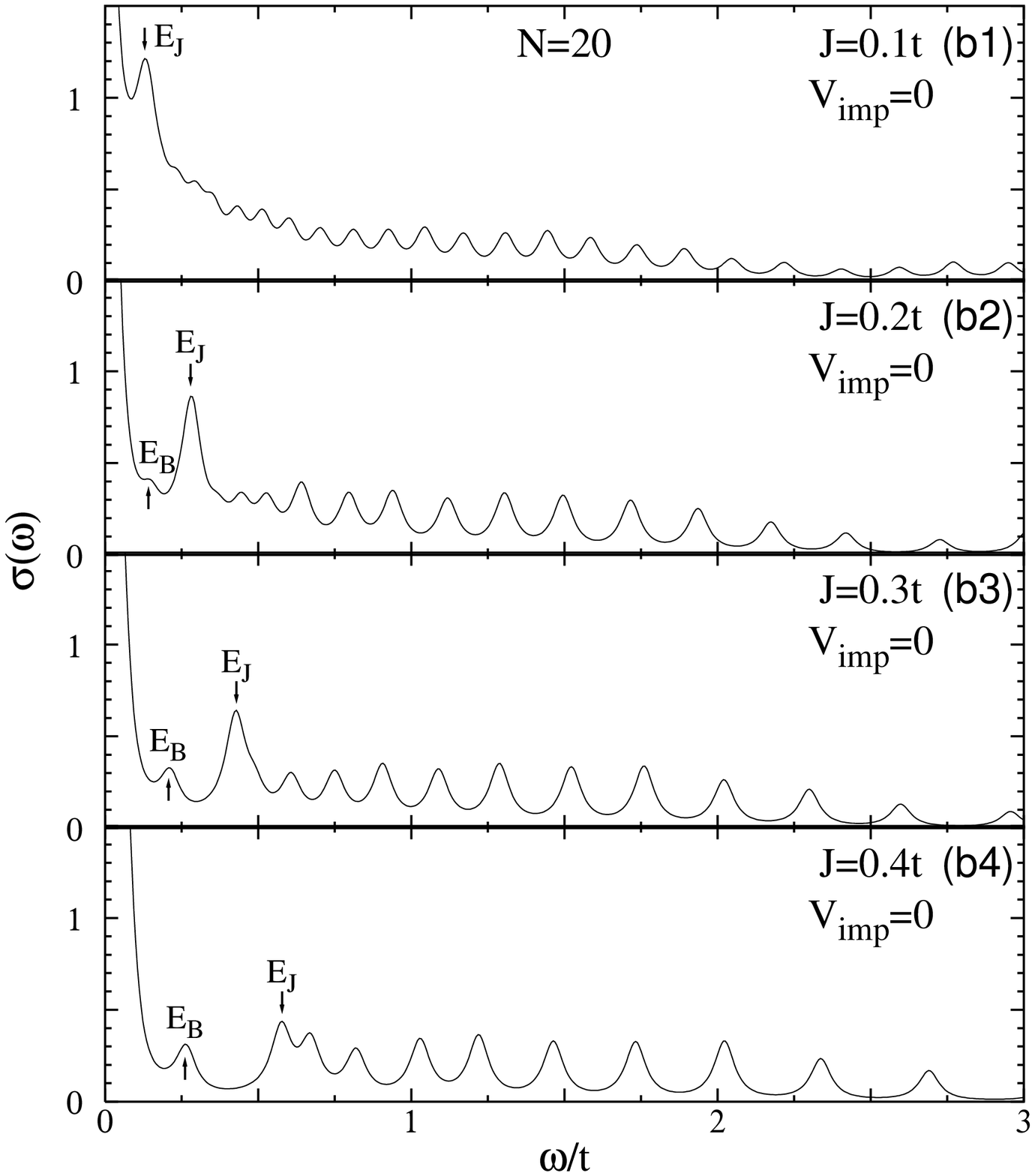, width=12cm} \\
(b)
\caption{}
\label{fig:opt16J}
\end{figure}

In Fig.~\ref{fig:opt16J}(a)
the predicted optical conductivity is shown
for various values of $J$ and $V_{\rm imp}=0$ 
on $4\times4$ square lattice.
A Drude peak is seen to appear in the zero-frequency limit 
$\omega \rightarrow 0$
for all chosen values of $J$.
For $J=0.1t$
a broad Drude peak is predicted with no special feature,
as is shown in Fig.~\ref{fig:opt16J}(a1).
As the Heisenberg interaction strength $J$ increases further,
interestingly enough, an additional small peak 
(denoted as $E_B$ and indicated by an upward arrow in the figure)
is predicted to occur at a low frequency,
as is shown in Figs.~\ref{fig:opt16J}(a2)--(a4).
In addition a large peak 
(denoted as $E_J$ and indicated by an downward arrow)
is seen to appear at a higher frequency.
This peak becomes increasingly separated 
from both the Drude peak and the small peak
with the increase of $J$.
The larger peak at a higher frequency may be directly associated with
the Heisenberg exchange correlation,
but not with the nonmagnetic impurity.

For further study of the low energy peak,
we choose $t\simeq 0.44 \mbox{~eV}$ for the hopping energy
and $J \simeq 0.128 \mbox{~eV}$ for the Heisenberg exchange energy
(i.e., $J \simeq 0.3t$)
obtained from a local-density-functional study. \cite{90Hybertsen}
Similar features without the disappearance of
the low energy peaks appear 
for the larger cluster of $\sqrt{20}\times\sqrt{20}$ size
as is shown in Fig.~\ref{fig:opt16J}(b).
The presence of the low energy peak $E_B$ may
not be subject to the finite size effect,
although quantitative differences may exist.

\begin{figure}[h]
\centering
\epsfig{file=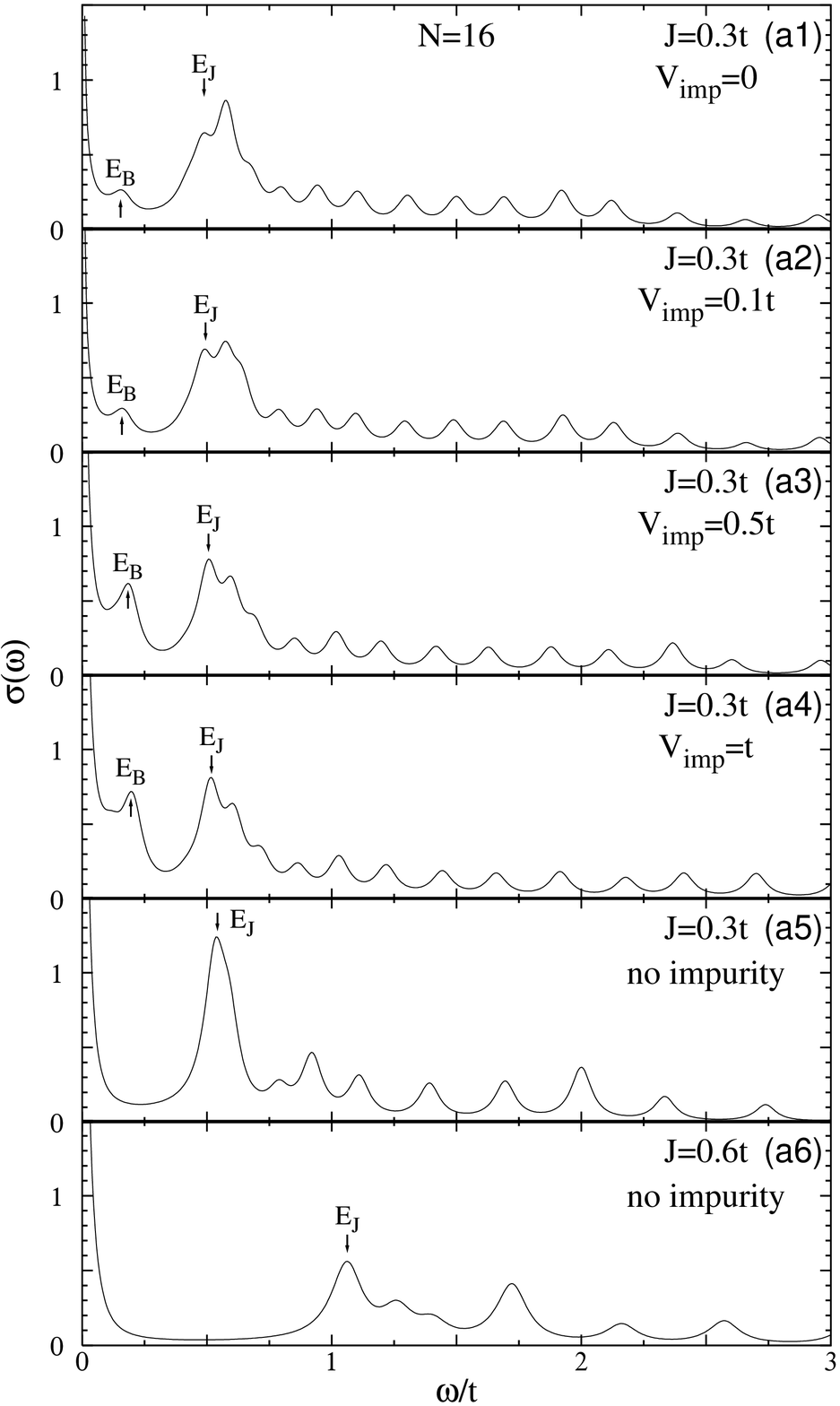, width=12cm} \\
(a) \\
FIG.~2. \\
\newpage
\epsfig{file=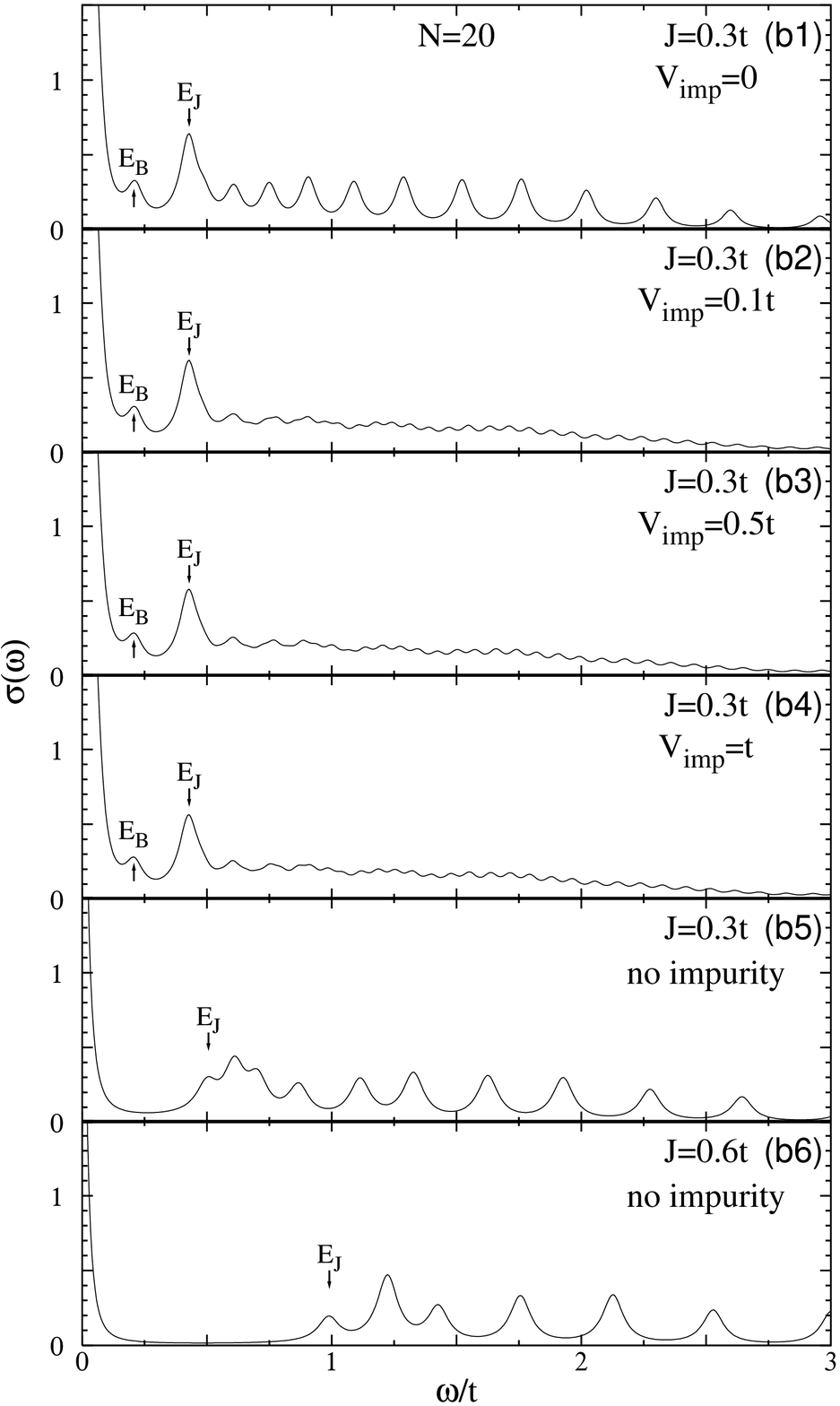, width=12cm} \\
(b) \\
\caption{}
\label{fig:opt16}
\end{figure}

In Fig.~\ref{fig:opt16}(a1)
the predicted optical conductivity is shown
for $J=0.3t$ and $V_{\rm imp}=0$.
The low-energy peak occurred at $E_B \sim 0.16t$.
As the strength of the impurity potential $V_{\rm imp}$ increases,
the intensity of the predicted peak becomes larger
and its position is seen to shift
to a higher frequency value of $E_B \sim 0.20t$.
This feature is shown in Figs.~\ref{fig:opt16}(a2)--(a4).
For comparison the optical conductivity 
in the absence of impurity
is displayed in Fig.~\ref{fig:opt16}(a5).
In the absence of impurity
the low energy peak disappeared,
while the position of the high energy peak $E_J$
remained unchanged at the same value of $J=0.3t$,
as is shown in Fig.~\ref{fig:opt16}(a4) and Fig.~\ref{fig:opt16}(a5).
The occurrence of the new small peak at a low-frequency region
is attributed to the presence of
the nonmagnetic impurity in the systems
of antiferromagnetically correlated electrons,
while the large peak at a high-frequency region
is contributed by the Heisenberg interaction between electrons.
Recent experimental data of optical conductivity
for ${\rm YBa_2}({\rm Cu}_{1-x}{\rm Zn}_x)_3{\rm O}_{7-\delta}$ crystals
exhibited a similar trend
(see the two figures in the second row of Fig.~2 in Ref.~11).
The observed optical conductivity for $4\%$ Zn-doped samples
measured at room temperature
showed a small hump near 750 cm$^{-1}$.
For the choice of $t\simeq 0.44 \mbox{~eV}$ [Ref.~13]
the predicted peak position at $E_B \sim 0.2t$
corresponds to the wave number of 710 cm$^{-1}$,
which is consistent with the measurement.
Based on the conjecture of Poilblanc {\it et al.}, \cite{94Poilblanc}
this small low energy peak may be the reflection of
forming a quasi-bound state as a result of resonant scattering 
with the nonmagnetic impurity.

As mentioned above,
the large peak which occurred at $E_J \sim 0.54t$
[Fig.~\ref{fig:opt16}(a5)] with $J=0.3t$
is attributed to the Heisenberg interaction between electrons.
This can be clearly understood from Fig.~\ref{fig:opt16}(a6):
in the absence of impurity 
the peak position is shifted to a higher frequency, $E_J \sim 1.1t$
for an increased value of $J$, say, $J=0.6t$.
Experimental studies of cuprate materials revealed that
broad bands associated with the Heisenberg interaction appear
near 0.2 eV (see Fig.~3 in Ref.~5).
Our predicted value of $E_J \sim 0.54t \simeq 0.24$ eV
is in good agreement with the experimental observation.

\section{Conclusion}
We have investigated the effect of nonmagnetic impurities
on the optical conductivity in the systems of
antiferromagnetically correlated electrons
by using the Lanczos exact diagonalization scheme.
It is found that 
a low-frequency peak in the optical conductivity
appears near 710 cm$^{-1}$ in the presence of nonmagnetic impurities,
in good agreement with experimental results.
The predicted low-frequency peak in optical conductivity
may be due to the resonant scattering of hole with the nonmagnetic impurity 
by allowing a possibility of forming a quasi-bound state.
In addition, a relatively high and broad peak is found to occur
at a high frequency region
as a consequence of the Heisenberg interaction between electrons,
in general agreement with observation in the peak position.

\newpage

\centerline{FIGURE CAPTIONS}
\begin{itemize}
\item[FIG. 1]
Optical conductivity versus frequency
for various values of $J$ and $V_{\rm imp}=0$.
Clusters of size (a) $4\times 4$ and
(b) $\sqrt{20}\times \sqrt{20}$ with one doped hole
in the presence of a single nonmagnetic impurity atom
are considered.
The $\delta$-functions are given with the width, $\epsilon=0.05t$.
$E_B$ indicates a low-energy peak owing to the presence of 
a nonmagnetic impurity with a positive charge
and $E_J$, a peak associated with Heisenberg exchange interaction.

\item[FIG. 2]
Optical conductivity versus frequency
for $J=0.3t$ and various values of $V_{\rm imp}$.
Clusters of size (a) $4\times 4$ and
(b) $\sqrt{20}\times \sqrt{20}$ with one doped hole
in the presence of one nonmagnetic impurity
are considered.
The $\delta$-functions are given with the width, $\epsilon=0.05t$.
$E_B$ indicates a low-energy peak owing to the impurity
and $E_J$ is associated with Heisenberg exchange correlation.
\end{itemize}

\newpage

\end{document}